\documentclass[letterpaper]{article}

\usepackage{natbib,alifeconf}  %% The order is important

\usepackage[dvipsnames,table,xcdraw]{xcolor}
\usepackage{graphicx}

\usepackage{adjustbox}
\usepackage[noEnd=true,commentColor=black]{algpseudocodex}
\usepackage{algorithm}
\usepackage{amsmath,amssymb,amsthm,amsfonts}
\usepackage{annotate-equations}
\usepackage{array}
\usepackage{bibunits}
\usepackage{bold-extra}
\usepackage{booktabs}
\usepackage{cancel}
\usepackage{caption}
\usepackage{circledsteps}
\usepackage{cite}
\usepackage{csvsimple}
\usepackage{etoolbox}
\usepackage{float}
\newfloat{algorithm}{t}{lop} % adapted from https://tex.stackexchange.com/a/6166/316176
\usepackage[margin=1in]{geometry}
\usepackage{hyperref}
\usepackage{import}
\usepackage{enumitem}
\usepackage{mathtools}
\usepackage{orcidlink}
\usepackage{physics}
\usepackage{placeins}
\usepackage{rotating}
\usepackage[subtle]{savetrees}
\usepackage{siunitx}
\usepackage{stix}
\usepackage{stmaryrd}
\usepackage{subcaption}
\usepackage{tabularx}
\usepackage{textcomp}
\usepackage{tikz}
\usetikzlibrary{arrows.meta}
\usepackage[most]{tcolorbox}
\usepackage{varwidth}
\usepackage{xparse}
\usepackage{pythonhighlight}
\usepackage{empheq}
\usepackage{cleveref}

\definecolor{seabornBrown}{RGB}{134, 86, 75}
\definecolor{seabornPink}{RGB}{255, 119, 193}
\definecolor{seabornPinkDark}{RGB}{152, 34, 116}

\newcolumntype{Y}{>{\centering\arraybackslash}X}

\newtcolorbox{mybox}{
enhanced,
boxrule=0pt,frame hidden,
borderline west={4pt}{0pt}{green!50!black},
colback=green!30!gray!15,
sharp corners,
parbox=false
}

\makeatletter
\def\mathcolor#1#{\@mathcolor{#1}}
\def\@mathcolor#1#2#3{%
  \protect\leavevmode
  \begingroup\color#1{#2}#3\endgroup
}
\makeatother

\makeatletter
\DeclareRobustCommand{\ctau}{{% capital tau
  \mathpalette\cap@greek\tau
}}
\DeclareRobustCommand{\csigma}{{% capital sigma
  \mathpalette\cap@greek\sigma
}}
\newcommand{\cap@greek}[2]{%
  \begingroup
  \sbox\z@{$#1t$}% measure a capital letter in the current style
  \resizebox{!}{\ht\z@}{$\m@th#1#2$}% resize tau to match
  \endgroup
}
\makeatother

\let\mycheckmark\checkmark
\renewcommand{\checkmark}{\textcolor{ForestGreen}{\mycheckmark}}

\theoremstyle{definition}

\makeatletter
\@addtoreset{proofpart}{theorem}
\@addtoreset{proofpart}{lemma}
\@addtoreset{proofpart}{corollary}
\makeatother

\hypersetup{breaklinks=true}

\renewcommand{\eqnhighlightheight}{\mathstrut}

\newcommand*{\colorboxed}{}
\def\colorboxed#1#{%
  \colorboxedAux{#1}%
}
\newcommand*{\colorboxedAux}[3]{%
  \begingroup
    \setlength\fboxrule{1pt}
    \colorlet{cb@saved}{.}%
    \color#1{#2}%
    \boxed{%
      \color{cb@saved}%
      #3%
    }%
  \endgroup
}

\makeatletter
\renewcommand{\boxed}[1]{\text{\fboxsep=.2em\fbox{\m@th$\displaystyle#1$}}}
\makeatother

\let\mythealgorithm\thealgorithm

\makeatletter
\newlength{\comment@width}

\renewcommand{\Comment}[1]{%
  \sbox0{#1}% measure
  \ifdim\wd0>\comment@width
    \setlength{\comment@width}{\wd0}%
  \fi
  \ifcsname comment@\arabic{algorithm}@width\endcsname
    \algorithmiccomment{\makebox[\csname comment@\mythealgorithm @width\endcsname][l]{#1}}%
  \else
    \algorithmiccomment{#1}%
  \fi
}
\AtBeginEnvironment{algorithmic}{\setlength{\comment@width}{0pt}}
\AtEndEnvironment{algorithmic}{%
  \immediate\write\@auxout{%
    \string\algcommentwidth{\mythealgorithm}{\the\comment@width}%
  }%
}
\newcommand{\algcommentwidth}[2]{%
  \global\@namedef{comment@#1@width}{#2}%
}
\makeatother

\MakeRobust{\Call}
\makeatletter
\def\algbackskip{\hskip-\ALG@thistlm}
\makeatother

\lstnewenvironment{mypython}[1][]{\lstset{style=mypython,#1}}{}

\definecolor{darkred}{HTML}{E32B60}

\definecolor{codegreen}{rgb}{0,0.6,0}
\definecolor{codegray}{rgb}{0.5,0.5,0.5}
\definecolor{codepurple}{rgb}{0.58,0,0.82}
\definecolor{backcolour}{rgb}{0.95,0.95,0.92}

\lstdefinestyle{mystyle}{
    backgroundcolor=\color{backcolour},
    commentstyle=\color{codegreen},
    keywordstyle=\color{magenta},
    numberstyle=\tiny\color{codegray},
    stringstyle=\color{codepurple},
    basicstyle=\ttfamily\footnotesize,
    breakatwhitespace=false,
    breaklines=true,
    captionpos=t,
    keepspaces=true,
    numbers=left,
    numbersep=5pt,
    showspaces=false,
    showstringspaces=false,
    showtabs=false,
    tabsize=2
}

\DeclareCaptionFormat{listing}{\rule{\dimexpr\textwidth\relax}{0.4pt}\par\vskip1pt#1#2#3\par\vspace{-5pt}\rule{\dimexpr\textwidth\relax}{0.4pt}}
\usepackage{pdflscape,lipsum}
\usepackage{eso-pic,zref-user}
\newcounter{cntsideways}
\makeatletter
\AddToShipoutPictureBG{%
 \ifnum\zref@extractdefault{rotate\number\value{page}}{page}{0}=0
  \PLS@RemoveRotate
 \else
  \PLS@AddRotate{90}%
 \fi}

\newcommand\rotatesidewayslabel{\stepcounter{cntsideways}%
 \zlabel{tmp\thecntsideways}\zlabel{rotate\zref@extractdefault{tmp\thecntsideways}{page}{0}}}

\makeatletter
\let\pragma@iinput=\@iinput
\def\@iinput#1{\xdef\@pragmafile{#1}\pragma@iinput{#1} }
\def\@pragmafile{default}
\def\pragmaonce{%
   \csname pragma@\@pragmafile\endcsname
   \global\expandafter\let \csname pragma@\@pragmafile\endcsname =  
}
\makeatother

\ifdefined\mydraft
\mydraft
\fi

\defaultbibliography{bibl}
\defaultbibliographystyle{apalike-ejor}

\begin{document}

\title{Extending a Phylogeny-based Method for Detecting Signatures of Multi-level Selection for Applications in Artificial Life}

% Each submission will undergo a double‑blind review process. To this end, submissions should NOT contain any element that could reveal the identity of the authors (author names, affiliations, funding details and acknowledgments), and should use the third person to refer to previous work by the authors.

\author{
    Matthew Andres Moreno\orcidlink{0000-0003-4726-4479}$^{1,2,3,4,\dagger}$,
    Sanaz Hasanzadeh Fard\orcidlink{0009-0007-0807-1339}$^{5,7}$,
    Luis Zaman\orcidlink{0000-0001-6838-7385}$^{1,2,4}$, \and
    Emily Dolson\orcidlink{0000-0001-8616-4898}$^{5,6,7}$%
    \mbox{}\\
    $^1$Department of Ecology and Evolutionary Biology
    $^2$Center for the Study of Complex Systems\\
    $^3$Michigan Institute for Data and AI in Society
    $^4$University of Michigan, Ann Arbor, United States\\
    $^5$Department of Computer Science and Engineering
    $^6$Program in Ecology, Evolution, and Behavior\\
    $^7$Michigan State University, East Lansing, United States\\
    $^\dagger$\texttt{morenoma@umich.edu}
}

\maketitle

\begin{bibunit}

\begin{abstract}
Multilevel selection occurs when short-term individual-level reproductive interests conflict with longer-term group-level fitness effects.
Detecting and quantifying this phenomenon is key to understanding evolution of traits ranging from multicellularity to pathogen virulence.
Multilevel selection is particularly important in artificial life research due to its connection to major evolutionary transitions, a hallmark of open-ended evolution.
\citet{BonettiFranceschi2024phylogenetic} proposed to detect multilevel selection dynamics by screening for mutations that appear more often in a population than expected by chance (due to individual-level fitness benefits) but are ultimately associated with negative longer-term fitness outcomes (i.e., smaller, shorter-lived descendant clades).
Here, we use agent-based modeling with known ground truth to assess the efficacy of this approach.
To test these methods under challenging conditions broadly comparable to the original dataset explored by \citet{BonettiFranceschi2024phylogenetic}, we use an epidemiological framework to model multilevel selection in trade-offs between within-host growth rate and between-host transmissibility.
To achieve success on our \textit{in silico} data, we develop an alternate normalization procedure for identifying clade-level fitness effects.
We find the method to be sensitive in detecting genome sites under multilevel selection with 30\% effect sizes on fitness, but do not see sensitivity to smaller 10\% mutation effect sizes.
To test the robustness of this methodology, we conduct additional experiments incorporating extrinsic, time-varying environmental changes and adaptive turnover in population compositions, and find that screen performance remains generally consistent with baseline conditions.
This work represents a promising step towards rigorous generalizable quantification of multilevel selection effects.
\end{abstract}

\section{Introduction} \label{sec:introduction}

A major unsolved problem in open-ended evolution research is the development of general-purpose methods for identifying major evolutionary transitions in individuality \citep{dolson2019modes}. Major evolutionary transitions, where the unit that can most accurately be described as an ``individual'' shifts (e.g., from a single-celled organism to a multicelled organism or from a prokaryotic cell to a eukaryotic cell), are thought to be a hallmark of open-ended evolution \citep{taylor2016open}. They are also a topic of great interest in evolutionary biology \citep{maynardsmithMajorTransitionsEvolution1997, szathmaryMajorEvolutionaryTransitions2015}. Recognizing major transitions is challenging, however, because most approaches require pre-determining what constitutes an individual pre- and post-transition, meaning they cannot recognize truly organic and unexpected transitions.

A major force in bringing about major evolutionary transitions is multi-level selection, i.e., selection occurring at different levels of organization simultaneously. Major transitions are generally thought to occur when selection on a collection of units in concert becomes stronger than selection on the individual units \citep{michodReorganizationFitnessEvolutionary2003, ratcliffNascentLifeCycles2017}. Measuring multi-level selection, then, is a promising proxy for measuring major transitions. Multi-level selection is also valuable in its own right, as it can shape evolutionary dynamics even when it doesn't lead to a major evolutionary transition. For example, the virulence vs. transmission tradeoff is an instance of multi-level selection that plays a key role in determining the course of viral evolution \citep{coombsEvaluatingImportanceBetweenhost2007, alizonVirulenceEvolutionTradeoff2009} (but see \citep{acevedoVirulencedrivenTradeoffsDisease2019}); individual viral particles may extract a selfish benefit from harming the host more, but at the cost of making the infected person feel sicker, which reduces the chance that any of the viral particles in their body can transmit to a new host.

Here, we set out to develop a general-purpose technique for measuring multi-level selection in artificial life systems. Such a technique must make as few assumptions as possible about the system. One assumption that should be fairly safe to make is that any evolving system will involve some form of reproduction, where organisms give rise to new organisms. Thus, it should almost always be possible to build phylogenies (i.e., ancestry trees)\footnote{One possible exception is systems with sexual reproduction (crossover). However, in most cases, it is still possible to build asexual phylogenies for each gene in these systems.}. Indeed, other techniques for quantifying open-ended evolution have successfully used phylogenies as input data \citep{dolson2019modes}. Similarly, it is relatively safe to assume that most evolving artificial life systems have heritable genetic information that could, if necessary, be represented as a sequence. Thus, we propose to base our multi-level selection metric on the following two input data types: phylogenies and genetic sequences.

% Double check that this is the right Volz citation
Encouragingly, an approach to detecting multi-level selection based on this data has already been proposed and used to screen for multi-level selection in a real-world SARS-CoV-2 phylogeny \citep{BonettiFranceschi2024phylogenetic}.
However, it has yet to be tested on data with a known ground-truth multi-level selection dynamics or data from an \textit{in silico} model system.
Thus, our primary goal in this work is to make the necessary adjustments to ensure that this metric correctly detects multi-level selection in Artificial Life systems.

\section{Methods} \label{sec:methods}

\subsection{Signature of Multilevel Selection}

Multilevel selection is most evident when there are conflicting incentives between fitness at different levels of biological organization.
For example, in the evolution of multicellularity, cell-level increases in proliferation rate can reduce the capability of the multicellular organism as a whole to survive and reproduce (e.g., cancer).
%Analagous split-level selective dynamics can also emerge outside the context of cooperative higher-level multicellularity.
Similarly, in the context of host-parasite dynamics, traits can provide individual-level benefits by facilitating within-host proliferation while harming the capability of the strain to infect further hosts --- e.g., due to prematurely killing the host or reducing interaction with other potential hosts (i.e., increased virulence).
Importantly, the units at lower levels of biological organization nearly always have shorter lifespans than the collectives that they are part of (e.g., cells have shorter lifespans than multicellular organisms, individual bees have shorter lifespans than bee colonies, etc.). Consequently, multi-level selection can manifest as a trade-off between the short- and long-term success of a lineage.

% graphics source https://docs.google.com/presentation/d/111HE56Ow0miZApHPBa18eyPAw8cdslZyBCavoWMix8E
\begin{figure}

\begin{minipage}{0.5\linewidth}
\centering
\includegraphics[width=\linewidth]{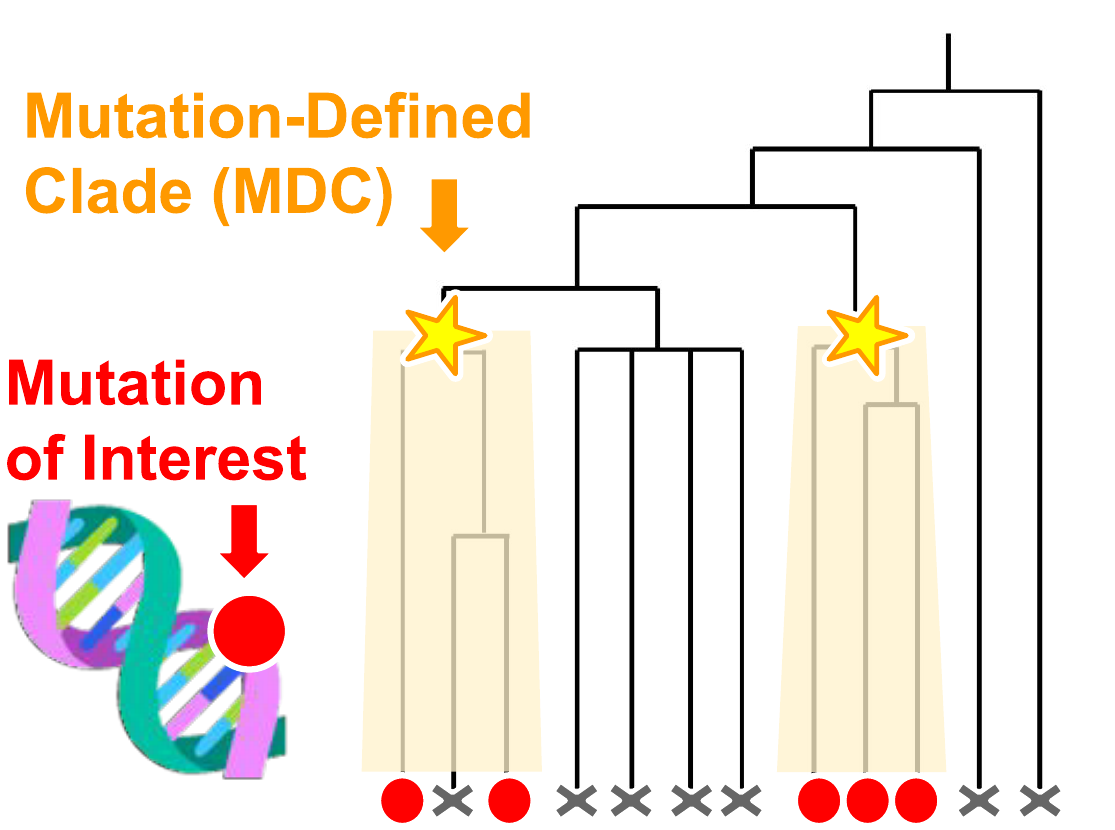}
\subcaption{mutation-defined clades}
\label{fig:mls-schematic:mdc}
\end{minipage}%
\begin{minipage}{0.5\linewidth}
\centering
\begin{minipage}{\linewidth}
\centering
\includegraphics[width=0.8\linewidth]{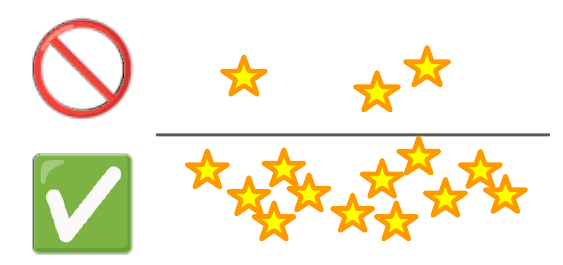}
\subcaption{mutation frequency}
\label{fig:mls-schematic:mutfreq}
\end{minipage}
\begin{minipage}{\linewidth}
\centering
\includegraphics[width=0.8\linewidth]{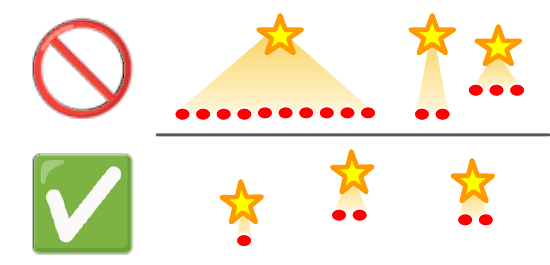}
\subcaption{fitness effect}
\label{fig:mls-schematic:fiteffect}
\end{minipage}
\end{minipage}

\caption{\textbf{Adapted method to screen for multilevel selection on an example mutation of interest.} a) The phylogeny for the mutation of interest. In this case, the mutation arises twice (indicated by stars), producing two mutation-defined clades (highlighted in yellow). Red circles indicate leaf taxa with the mutation. b) Mutations under MLS should be observed more often than expected by chance. c) Mutations under MLS should go on to produce smaller clades than expected by chance.}
\label{fig:mls-schematic}

\end{figure}

In the presence of multilevel selection dynamics, therefore, there exist mutations that can briefly surge in prevalence but ultimately be swept to extinction.
In this work, we seek to identify these specific genome sites.
We adopt the strategy proposed by \citep{BonettiFranceschi2024phylogenetic}, wherein these sites are identified as
\begin{enumerate}[noitemsep,topsep=0pt,parsep=0pt,partopsep=0pt]
\item appearing in the population at a higher rate than expected for a random neutral mutation (given transient short-term fitness benefit), and
\item being associated with smaller, shorter-lived clades than expected for neutral mutations (due to longer-term disadvantage).
\end{enumerate}

This approach, then, requires two types of information: (1) genome sequences, so that the sets of mutations carried by different individuals in a population can be identified, and (2) phylogenies, so that lineage history outcomes may be assessed.
In biological study systems, these two types of data are typically intrinsically linked because phylogenies are inferred from genome sequences.

In digital study systems, phylogeny is usually tracked directly and the exact point at which a mutation originated can be recorded.
However, direct tracking is impractical in some scenarios, like biological model systems or very large-scale digital simulations \citep{ackley2023robust,moreno2024trackable}. In these systems, much as in biology, the point in evolutionary history where mutations actually arose must be estimated.
\citet{BonettiFranceschi2024phylogenetic} use an \textit{ad hoc} approach to accomplish this estimation. They label tips (i.e., extant taxa) according to whether or not they contain a given mutation and identify inner nodes for which $\geq75\%$ of descendant tips on one daughter branch contain a mutation but $<75\%$ of nodes on the other do not contain that mutation.
The clade of individuals that share a mutation through common descent is termed as a \textit{Mutation-Defined Clade (MDC)}.
In this work, we use direct tracking approaches and the term MDC instead refers to the exact point at which a mutation arises.

In sum, therefore, our goal is to identify particular genome sites for which (1) a higher quantity of MDCs are observed than expected by chance and (2) when MDCs do occur, they tend to be smaller and/or more short-lived than would be expected by chance.
(Figure \ref{fig:mls-schematic}).

In the next section, we will describe the original statistical algorithm proposed by \citet{BonettiFranceschi2024phylogenetic} for this purpose, and our proposed modifications.

\subsection{Fitness Effect Screen Methodology}

One of the challenges of characterizing fitness outcomes from a phylogeny is accounting for confounding factors contributing to clade fates.
Several possible examples include: further clade history after the point at which a clade is sampled (e.g., recency bias), differences in genetic background having effects on selection (e.g., emergence of high-transmission SARS-CoV-2 strains like alpha, beta, omicron, etc.), and time-varying environmental effects (e.g., habitat/resource availability, predation).

\citet{BonettiFranceschi2024phylogenetic} propose an elegant approach to account for these issues: compare each MDC and its nearest non-ancestral relative (its ``sister'' clade).
% These clades and,
Essentially, these clades can be interpreted as differing solely in terms of the putatively causal influence of the mutation of interest.

% graphics source https://docs.google.com/presentation/d/111HE56Ow0miZApHPBa18eyPAw8cdslZyBCavoWMix8E
\begin{figure}[htb]
\centering
\begin{minipage}{0.47\linewidth}
\centering
\includegraphics[width=0.7\linewidth]{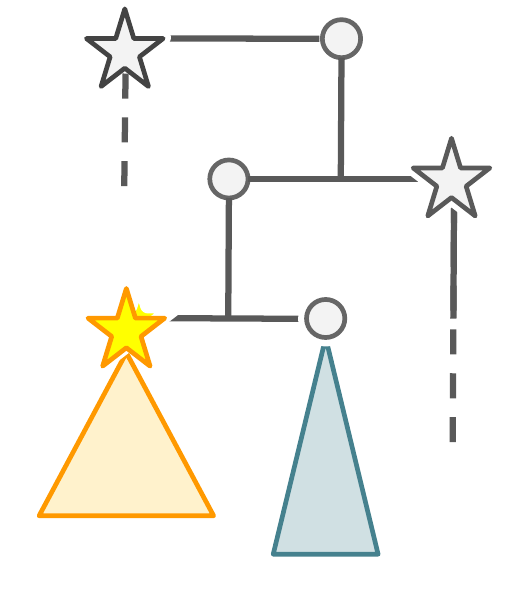}
\subcaption{one sister clade comparator}
\label{fig:screen-strategy:tree-balance}
\end{minipage}
~~
\begin{minipage}{0.47\linewidth}
\centering
\includegraphics[width=0.7\linewidth]{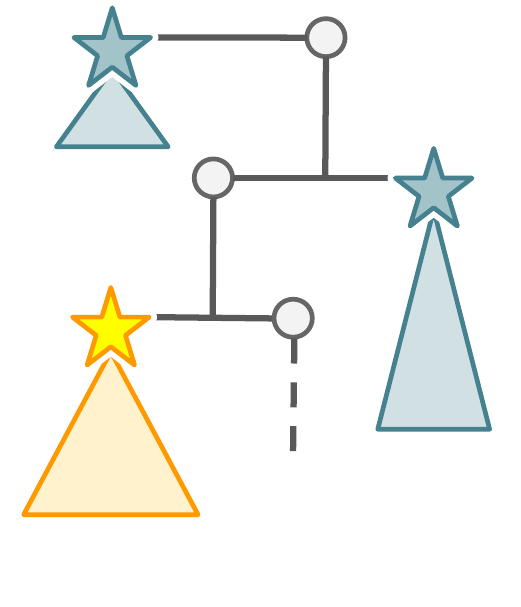}
\subcaption{multiple MDC comparators}
\label{fig:screen-strategy:window}
\end{minipage}
\vspace{-1ex}
\caption{%
\textbf{Fitness-effect screen strategies.}
\footnotesize
Yellow star indicates mutation of interest, yellow triangle indicates the clade it defines. Teal triangles indicate comparator clades. Teal stars indicate mutations defining comparator clades. Grey stars indicate mutations defining clades that are neither the focal clade nor comparator clades.
a) As proposed by \citep{BonettiFranceschi2024phylogenetic}, the focal clade is compared to its sister clade (which may or may not be mutation-defined), b) The focal clade is compared to a collection of other mutation-defined clades (MDCs).
}
\label{fig:screen-strategy}
\end{figure}

However, in adapting this approach to our \textit{in silico} study system, we ran into some challenges (described below) that led us to develop an alternate normalization strategy. Instead of just comparing clades to their sister clades, we identified clades' rankings within a broader set of selected comparator clades.
Specifically, we normalize against the pooled set of MDCs for \textit{other} mutations.
Figure \ref{fig:screen-strategy} provides a schematic comparison between the sister-clade and MDC comparator approaches.

Our motivation in diverging to develop the proposed comparator-based approach is twofold, driven by the unique characteristics of phylogenetic data from digital study systems.
First, we sought to enhance statistical power of the assay.
%TODO could show results with very small comparator groups to back up this point
Power is of particular concern for \textit{in silico} agent-based model systems, due to their typically smaller populations relative to microbial and viral populations.
In particular, under the sister clade comparator approach, the statistics measured for each individual MDC are strongly influenced by the status of the comparator clade.

Another challenge associated with densely-sampled, exactly-tracked phylogenies is complicating artifacts in tree topology.
For instance, high density of short tips descended from internal nodes\footnote{These tips are the result of one-off mutations that never reach a large population size. They are common in exactly tracked phylogenies, but rare in phylogenies reconstructed from real-world samples, as the odds of sampling these tips are low.} can introduce a bias towards comparison against these internal tips as sister clades.
Likewise, the existence of true multifurcations, which must be arbitrarily resolved in order to perform binary comparisons, further complicates matters \citep{dolson2024phylotrack}.
Relatedly, another useful factor of the comparator-based approach is in normalizing exclusively against other phylogeny nodes where mutations actually arose, thus balancing out possible bias in node characteristics where MDCs are observed (e.g., due to factors such as incubation time, infection order, phylogeny artifacts, etc.).
In contrast, the sister comparator approach includes comparison against non-MDC nodes.

\citet{BonettiFranceschi2024phylogenetic} apply a combination of three clade statistics to quantify empirical fitness characteristics of clades: (1) clade size, (2) clade duration, and (3) clade growth rate.
The first two properties are, respectively, measured as 1) the ratio between the number of leaves in a clade and the number in its sister clade and 2) the ratio of its lifetime (elapsed time between the estimated MDC-originating event and the extinction of its last leaf) and that of its sister.

The third metric has a more subtle calculation and interpretation.
\citet{Volz2023} demonstrates that the relative growth rate between two clades can be estimated by performing binary logistic regression.
In this approach, membership in clade A versus clade B is considered as a binary response variable, and the predictor variable is taken as chronological time.%
\footnote{%
Logistic regressions are moderately computationally expensive to perform,
In anecdotal testing, we found directionally-consistent results with orders-of-magnitude faster runtime could be achieved by applying recently-proposed fast approximations for binary logistic regression \citep{Saran2025}.
Further investigation is warranted in this regard.
}

Due to the multiple-comparator approach pursued in this work, we generalize clade fitness measures beyond the binary measures used by \citet{BonettiFranceschi2024phylogenetic}.
For the clade size and clade duration measures, we normalize against the comparator group through a simple percentile ranking.
(Because of strong discretization effects, we take a mean rank measure for tie-breaking.)
For the sake of simplicity, we omit the logistic clade growth measure from current work, given that devising a generalization beyond simple binary comparison would be more involved.

% graphics source https://docs.google.com/presentation/d/111HE56Ow0miZApHPBa18eyPAw8cdslZyBCavoWMix8E
\begin{figure}
\centering
\begin{minipage}{0.47\linewidth}
\centering
\includegraphics[height=0.75\linewidth,angle=270]{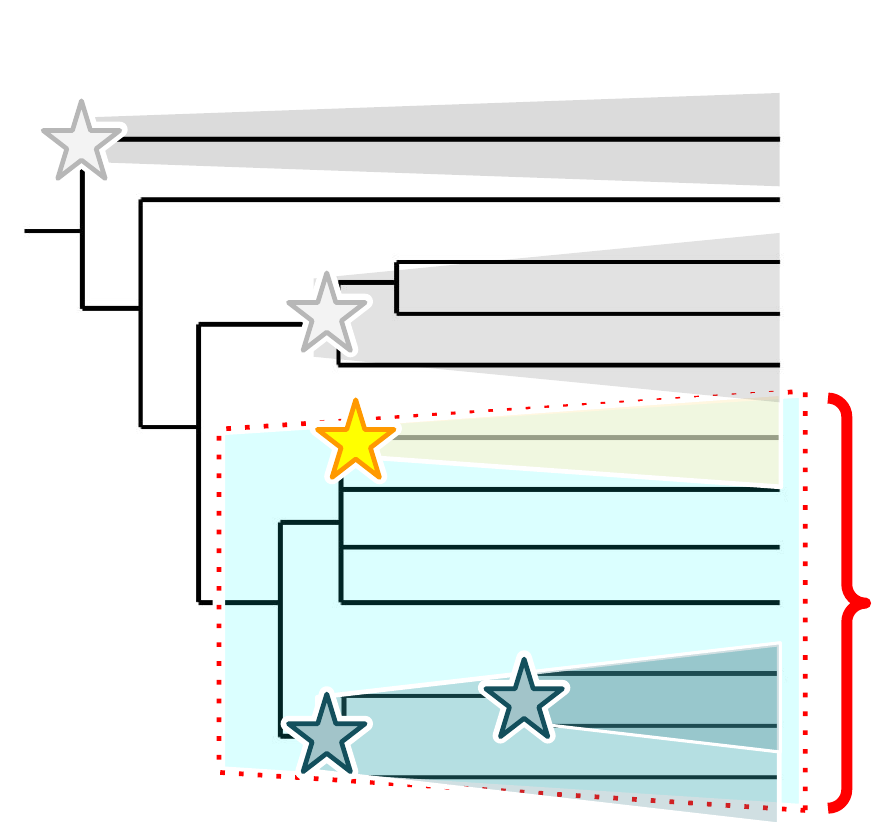}
\subcaption{phylogeny-based}
\label{fig:comparator-selection:clade}
\end{minipage}
~~
\begin{minipage}{0.47\linewidth}
\centering
\includegraphics[height=0.75\linewidth,angle=270]{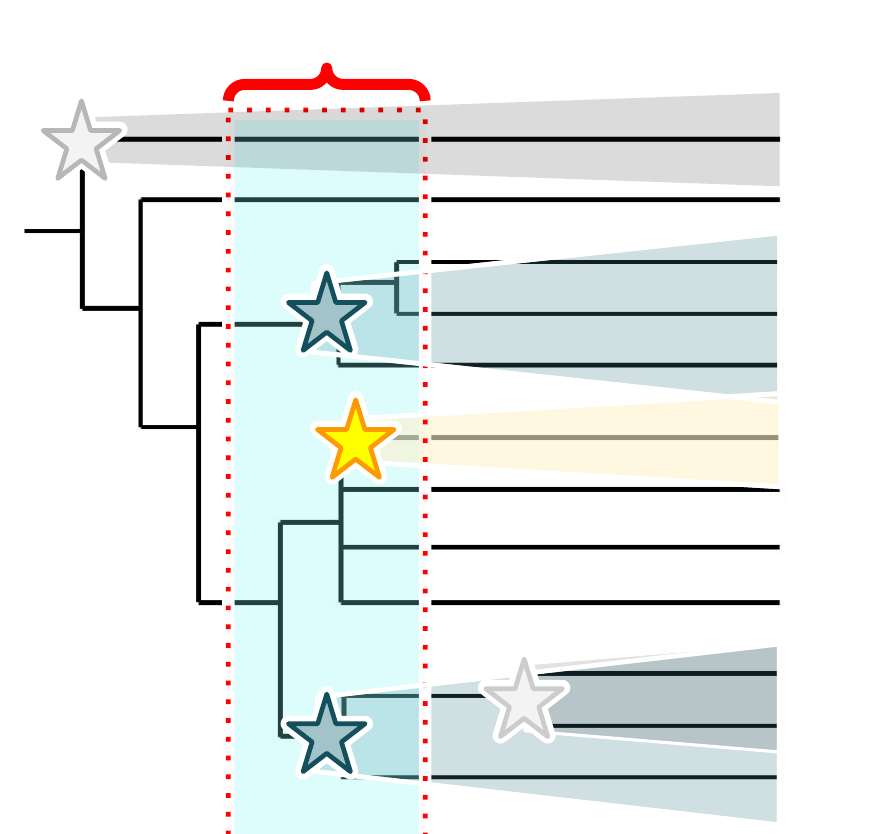}
\subcaption{epoch-based}
\label{fig:comparator-selection:window}
\end{minipage}

\begin{minipage}{\linewidth}
\caption{%
\textbf{Strategies for comparator selection.}
\footnotesize Yellow star indicates mutation of interest, teal star indicates mutations defining comparator clades. Grey stars indicate mutations defining clades that are neither the focal clade nor comparator clades.
a) Comparator group is all mutation-defined subclades of a larger clade. b) Comparator group is all mutation-defined clades that originated within a specific timespan.
}
\label{fig:comparator-selection}
\end{minipage}
\vspace{-3ex}
\end{figure}

To account for confounding and biasing factors mentioned above, we fine-tune the set of comparator MDCs sampled to provide a more apples-to-apples comparison.
Figure \ref{fig:comparator-selection} summarizes the two comparator selection methods explored in this work.
To account for variation due to recency-bias and time-varying environmental factors, the tree may be partitioned into discrete temporal ``epochs,'' and comparison limited to other MDCs originating within the same epoch.
To account for variation on account of extrinsic genetic factors, the tree may be partitioned into a set of comparator clades.
These may be selected using domain knowledge (in this work, phylogenetic partitions are performed based on WHO strain variant).
In future work, it would also be promising to explore conducting these comparisons on the basis of partitions automatically inferred from aspects of tree topology, an area of emerging phylostatistical methodology \citep{Lefrancq2025}.

To perform our statistical tests for deviation from expected values, we perform a bootstrap test to determine whether the mean percentile ranking of the focal clade within the comparator group is significantly different from the 50th percentile (as we would expect by chance).
% https://github.com/mmore500/multilevel-selection-concept/blob/6c77a8afbeb760cd8fa878c6fec8fbd8766aaddb/pylib/_calc_compsummary_stats.py
We used a bootstrap sample size of 1000,000.
As a caveat, bootstrap testing is known to be unreliable for small sample sizes (e.g., $\sim30$ observations \citep{Efron1994}); thus, where reporting bootstrap results, we indicate where this condition is violated.
Another important subtlety of our normalization approach is that comparator groups should be comprised of non-overlapping partitions.
Under this condition, sampling is guaranteed to be evenly distributed across the percentile scale (because the percentile scale is consistently defined in terms of sampled values) --- ensuring the validity of comparing against null expectation of the 50th percentile as the group mean.

To account for intrinsic uncertainty in real-world biological data,
\citet{BonettiFranceschi2024phylogenetic} take several further quality-control steps by excluding the following from consideration: 1) data beyond a fixed minimum clade size threshold, and 2) genome regions known to be associated with sequencing anomalies.
However --- given that we are using perfect-fidelity data from a digital model --- for the sake of simplicity and to maximize statistical power of our datasets, we omit these adjustments.

\subsection{Mutation Frequency Screen Methodology}

A variety of statistical tests exist to identify elevated mutation frequency \citep{mcdonaldAdaptiveProteinEvolution1991, narumComparisonFSTOutlier2011, yangStatisticalPropertiesBranchSite2011}.
As we found that the primary challenge in detecting multilevel selection effects is in testing for group-level fitness effects (e.g., transmission efficacy), we adopted a simple approach to screening mutation frequency for the purposes of present work.

To begin, we performed a simple calculation of baseline mutation frequency $\lambda$ as the mean number of net observed non-focal MDCs per instantiated non-focal background site.
To assess the significance of the observed mutation count's deviation from this expectation, we applied the Poisson survival function to calculate the probability of observing an equivalent or more extreme mutation count under the assumption of mutations being distributed $\sim \mathsf{Pois}(\lambda)$.

\subsection{Covasim Simulation Platform}

% graphics source https://docs.google.com/presentation/d/111HE56Ow0miZApHPBa18eyPAw8cdslZyBCavoWMix8E
\begin{figure*}[htbp]
\begin{minipage}{0.65\linewidth}
\includegraphics[width=\linewidth]{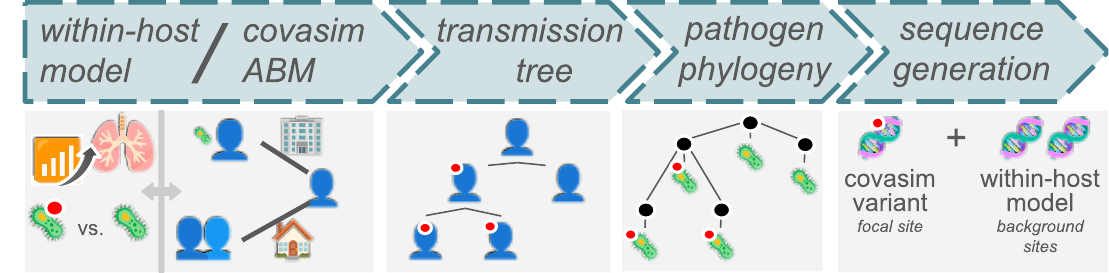}
\end{minipage}
\begin{minipage}{0.34\linewidth}
\caption{\textbf{Test data generation pipeline.} \footnotesize Joint within-host/between-host evo-epidemiological modeling produces a sample transmission tree, then converted to a plausible ground-truth pathogen phylogeny. Accompanying strain metadata determines genetic sequence content for each viral particle.}
\label{fig:covasim-schematic}
\end{minipage}
\vspace{-3ex}
\end{figure*}

We chose to work with an epidemiological model system for three reasons: (1) to leverage the stochastic dynamics and strong temporal fluctuations inherent among evolutionary dynamics of pathogens to create a challenging testbed for this statistical methodology, (2) due to the important practical implications of understanding the role of multilevel selection in the evolution of virulence, and (3) for comparability with original work by \citet{BonettiFranceschi2024phylogenetic}, who applied their analysis to a 1.2 million-sequence Covid-19 dataset obtained from the COVID-19 Genomics UK (COG-UK) Consortium \citep{marjanovic2022covid}.

\subsubsection{Covasim Platform}

We leveraged a pre-existing agent-based epidemiological modeling platform called Covasim for this work.
This software includes features to capture a variety of aspects of relevant real-world factors, including the structure of social networks within a population (e.g., school, work), development of host immunity, and public health responses (quarantine policy, community containment efforts, etc.).

For our experiments, we used population sizes of either 200,000 or 1.2 million agents.
% https://github.com/mmore500/multilevel-selection-concept/blob/6c77a8afbeb760cd8fa878c6fec8fbd8766aaddb/pylib/_make_cv_sim_uk.py
We simulated a 650-day window between Jan 1, 2020 and Oct 31, 2021.
Five replicate simulation trials were conducted per observation, except for some large-scale simulation trials for which we only ran three replicates.

For the most part, we used default Covasim parameter settings, although we applied the fast-waning immunity configuration described in Covasim documentation to prevent extinctions of viral populations.
For simulations configured to test robustness to environmental variability, we used an existing suite of configurations emulating the timeline of vaccination rollouts and public health policies within the United Kingdom \citep{Panovska2025Covid19Analysis}.
Experiments were parameterized to model the original Wuhan Covid-19 strain, except for multi-strain experiments, which additionally imported the alpha, beta, and gamma strains at the approximate dates where they were first observed in the course of the Covid-19 pandemic.
Due to practical limitations, we used a baseline population size of 200,000 agents in most cases.
However, to assess the effect of dataset size on detection sensitivity, we replicated some experiments with a larger population size of 1.2 million agents.

Although the Covasim framework included capability to represent a fixed number of predefined strains arising at predetermined timepoints, in order to model the \textit{in situ} evolution of genetic traits we had to extend the existing code with a within-host evolution model, which we describe next.

\subsubsection{Within-host Model}

We used an aggregate-based model to track the composition of within-host viral populations for each member of the Covasim population.
For this purpose, we inspected Covasim state between one-day update steps to identify all new infections.
Within each infected host, we seeded an aggregate count for copies of the infecting genotype --- inferred from Covasim's strain label for the infection.
Then, as day cycles elapsed in the primary model, aggregate genotype counts grew through a doubling process up to a fixed within-host carrying capacity.
With each doubling, counts of mutants to other possible genotypes were drawn from a binomial distribution.
These mutant counts were subtracted from their source genotype aggregate and added to their destination genotype aggregate.
In this manner, we sought to capture Luria-Delbruck dynamics where mutants that arose by chance early in an infection had a greater impact on population composition than mutants arising later on \citep{luriaMutationsBacteriaVirus1943}.
To tie within-host dynamics back to the Covasim model of host-to-host transmission, we sampled a genotype at random from each active within-host population and wrote its corresponding strain identifier back to Covasim's ``active strain'' identifier for its host.
This process was performed between each daily update cycle, allowing the genetic composition of transmission events to fluctuate stochastically over time based on the composition of the within-host population.

We parameterized the per-replication mutation rate, within-host carrying capacity, and doubling rate of within-host viral growth as $2\times10^{-6}$ probability per site, $10^{10}$ viral particles, and 4 hours, respectively, based on reference figures from empirical literature \citep{Markov2023,Sender2021,Gunawardana2022}.

In our experiments, we defined a ``focal'' genome site, which was used as the target for testing screening methods.
However, evaluations also required additional neutral ``background'' genome sites in order to assess the capabilities of screens to selectively identify the properties of the focal site.
In order to ensure statistical interchangeability between focal sites and background sites, we used additional instantiations of the aggregate within-host model to simulate the genetic composition of the background sites; the process by which mutations in background sites arose was therefore identical to that for fitness-neutral configurations of the focal site.

One limitation of this approach for generating background mutation data is that it does not capture linkage effects between mutations on the time scale of within-host evolution, as the genetic composition of each site fluctuates independently during within-host growth in our model.
We also do not capture various nuances of true nucleotide sequence data, such as site-specific biases in mutational rates or outcomes.
A final limitation is in the number of background sites simulated.
Due to practical limitations, for our experiments, it was feasible to simulate 50 total sites for smaller-scale 200,000-agent trials and 20 total sites for larger-scale 1.2 million-agent simulations.
Although higher site counts inherently increase the amount of sensitivity and selectivity necessary to reliably detect sites under multilevel selection, we expect that more sites would support a richer corpus of comparator MDCs, which in turn would enable finer-grained normalization of clade fitness statistics.

\subsubsection{Phylogeny Tracking}

We used Covasim's infection log feature as the basis for collecting phylogenetic data from our experiments.
Thus, our resolution in tracking pathogen ancestry was limited to the resolution of events where the pathogen transmitted between hosts.
We did not conduct supplemental within-host phylogeny tracking, as virus particle-level ancestry would be poorly defined in our aggregate-based within-host model.
However, we did record strain/genotype on the basis of a sampled strain/genotype from the within-host population at day 8 of infection, rather than the founding strain, to better reflect the character of data generated in real-world sampling (because a sample is unlikely to be taken from a real-world patient immediately upon infection).

To adjust the topology of our synthetic dataset to more closely match the characteristics of biological data, we interpreted all sampled genomes as tip nodes, even if they were an inner node in the transmission tree.
We accomplished this goal by applying a transform where all inner nodes were copied into their own child list as an additional single-taxon tip branch.
In future work, it will be important to explore the impact of limitations in phylogeny resolution and reliability on the performance of MLS detection methods.

\subsection{Treatment Conditions}

To assess the capability of evaluated methodology in discriminating the signature of multilevel selection effects from other scenarios, we compared screening outcomes across a set of three ground-truth configurations for the fitness effect of our focal genome site:
\begin{enumerate}[noitemsep,topsep=0pt,parsep=0pt,partopsep=0pt]
\item \textbf{Sben/Gdel}: true multilevel selection dynamics,
\item \textbf{Sben/Gneu}: within-host growth rate advantage,
\item \textbf{Sneu/Gneu}: neutral dynamics,
\end{enumerate}

Under ``Sben'' (``self-beneficial'') conditions, mutations to the focal site confer an advantage to within-host growth rate.
Under ``Gdel'' conditions (``group deleterious''), mutations to the focal site confer a penalty to the between-host transmission rate in the Covasim simulation layer.
For our non-neutral treatments, we considered a spread of three effect sizes: 1.1x, 1.3x, and 2x.

Given our objective to assess the capability of the proposed method to detect multilevel selection tradeoffs, we did not include ``Gben'' (group beneficial) focal site properties within our survey, as these are easier to differentiate compared to Gneu conditions and typically leave well-understood signatures of adaptive sweeps within phylogenetic history.
We did, however, assess the efficacy of multilevel selection detection within a broader context of adaptive evolution, via a multi-strain Covasim configuration.

% \subsubsection{Robustness Assessment}

%In our experiments, we additionally sought to assess the robustness of multilevel selection detection across a set of possible complicating factors.
%Under the ``vanilla'' baseline configuration of Covasim used for our primary set of experiments, social network structure and agent behavior remained consistent across simulated time (although population-level immunity levels did fluctuate as population members encountered infections), and adaptive evolution of transmission characteristics did not occur.

% To assess the effect of dataset size on detection sensitivity, we also replicated some experiments with larger population sizes of 1.2 million agents (versus the baseline population size of 200,000 agents).

% In further work to assess the efficacy, of
% \noindent Complicating factors not considered:
% \begin{itemize}
%  \item sampling bias
%  \item pleiotropy/epistasis
%  \item virulence (not considered/modeled directly)
% \end{itemize}

We report details on average case counts and focal MDC/mutant counts across surveyed treatment conditions in our supplementary material \citep{ExtendingPhylogenybasedMethod2025}.
Notably, simulation conditions incorporating public health responses generated substantially smaller pathogen phylogeny datasets, on account of reduced viral prevalence.

\subsection{Software and Data Availability} \label{sec:materials}

Software materials for this work are available via GitHub at \href{https://github.com/mmore500/multilevel-selection-concept}{\texttt{mmore500/multilevel-selection-concept}}. \citep{matthew_andres_moreno_2025_15549530}.
Data and supplemental materials are available via the Open Science Framework at \url{https://osf.io/37fv8} \citep{ExtendingPhylogenybasedMethod2025,foster2017open}.

This project uses data formats and tools associated with the ALife Data Standards project \citep{lalejini2019data} and benefited significantly from open-source scientific software \citep{2020SciPy-NMeth,harris2020array,reback2020pandas,mckinney-proc-scipy-2010,waskom2021seaborn,hunter2007matplotlib,moreno2023teeplot,moreno2022hstrat,kerr2021covasim,dolson2024phylotrack,franceschi2024mlscluster,huertacepas2016ete3,moreno2024apc,lam2015numba,moreno2024dendropy,cock2009biopython,demaio2022phastsim}.

% Sequence data from NCBI Virus \citep{brister2014ncbi}: wildtype \citep{wu2020new,MN908947.3}, OY320691.1 \citep{OY320691.1}, OM095211.1 \citep{OM095211.1}, OY317474.1 \citep{OY317474.1}, OY324687.1 \citep{OY324687.1}, OQ430688.1 \citep{OQ430688.1}.

\section{Results and Discussion} \label{sec:results}

In this section, we report results of trials testing the efficacy of proposed extended methods for screening for multilevel selection effects against ground-truth data generated \textit{in silico}.
We discuss the fitness effect screen component performance first, followed by discussion of the mutation frequency screen component.

\subsection{Example Fitness Effect Screen}

\begin{figure}

\centering
\includegraphics[width=0.85\linewidth]{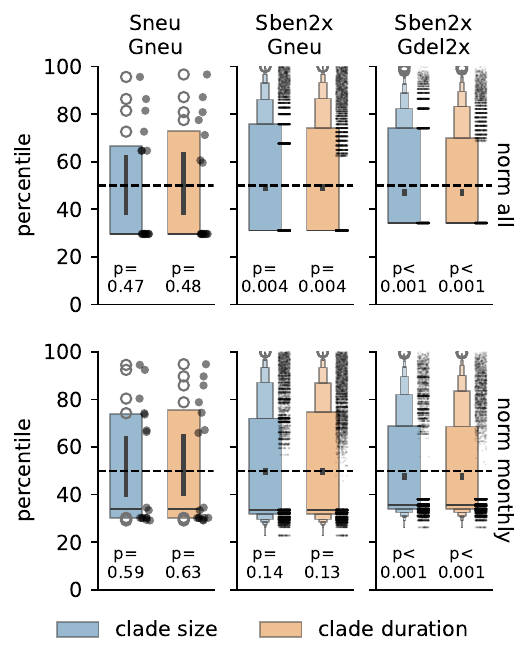}
\vspace{-1ex}
\caption{%
\textbf{Sample fitness effect screen result, baseline conditions.}
\footnotesize
Boxen plots represent distributions of percentile-normalized clade fitness metric values for focal site MDCs, with strip plot dots corresponding to individual MDCs.
Vertical bars indicate bootstrap 95\% confidence interval; annotated $p$ values correspond to bootstrap comparison of sample mean against null expectation of mean equal to 50th percentile.
Bottom row shows results for MDC comparator sets partitioned by simulation month; top row comparator sets used for percentile ranking included all observed MDCs.
Sben/Gdel treatment corresponds to true multilevel selection dynamics, tested at strong ($2\times$) effect size.
}
\label{fig:results-vanilla-example}
\vspace{-2ex}
\end{figure}

As a first step in evaluating our comparator-based approach for assessing MDC fitness outcomes, we compared the results of sample assays for deleterious between-host transmission effects across true MLS (Sben/Gdel) and control (Sneu/Gneu and Sben/Gneu) treatments.
For illustrative purposes, we show as examples the strongest-surveyed 2$\times$ magnitude for Gdel and Sben effects.

Figure \ref{fig:results-vanilla-example} compares the distribution percentile-normalized clade fitness statistics for focal MDCs across the neutral, within-host advantage, and true multilevel selection tradeoff treatments.
For multilevel selection Sben/Gdel conditions, bootstrap testing indicates the mean percentile ranking of MDC clade sizes and durations to both be significantly below the 50th percentile (%$\mu=TODO$,
$p<0.001$;
%$\mu=TODO$,
$p<0.001$).

Note that MDC fitness effects are falsely detected under Sben/Gneu conditions where comparator normalization included all MDCs (%$\mu=TODO$,
$p=0.004$;
%$\mu=TODO$,
$p=0.004$).
Indeed, we found the naive approach where every MDC was included in the comparator set for all other MDCs to generally produce a high false-positive rate across Sneu/Gneu and Sben/Gneu tests.
However, as seen in this example, we found that binning comparator sets by simulation month was effective in improving test specificity, and so adopted the approach of partitioning comparator sets by month for all subsequently-discussed results.

For Sneu/Gneu conditions shown in Figure \ref{fig:results-vanilla-example}, note that too few mutations were observed for the bootstrap test to be considered reliable.

\subsection{Fitness Effect Screen}

\begin{figure*}
\begin{minipage}[t]{0.65\linewidth}
\vspace{0pt}
\includegraphics[
    height=1.73in,
trim={0 0 0.95cm 0}, clip]{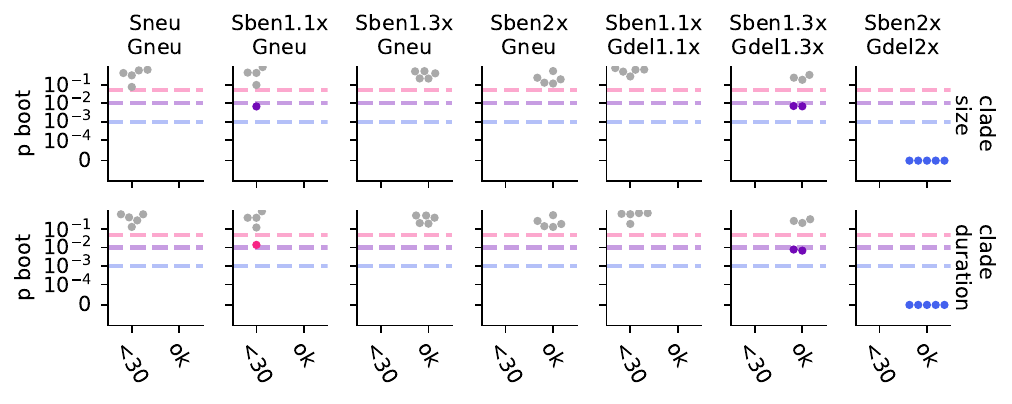}

\end{minipage}%
\begin{minipage}[t]{0.35\linewidth}
\vspace{0pt}
\includegraphics[
    height=1.652in,
    % width=0.497\linewidth,
trim={0.6cm 0 0.92cm 0}, clip]{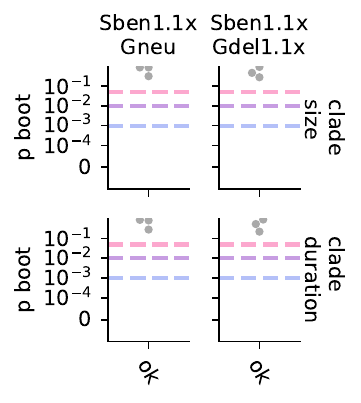}
\includegraphics[
    height=1.652in,
    % width=0.483\linewidth,
trim={1.65cm 0 0 0}, clip]{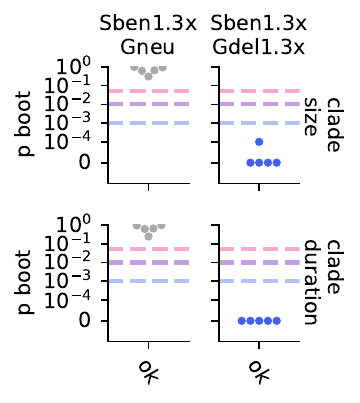}
\end{minipage}

\vspace{-2.4ex}

\begin{minipage}{0.65\linewidth}
\subcaption{host population size 200,000}
\label{fig:vanilla-summary-results:fit-effect-small}
\end{minipage}%
\begin{minipage}{0.35\linewidth}
\subcaption{host population size 1.2 million}
\label{fig:vanilla-summary-results:fit-effect-big}
\end{minipage}

\vspace{-1ex}

\caption{%
\textbf{Fitness effect screen results, baseline conditions.}
\footnotesize
Dots correspond to screen results for focal site in one replicate trial.
Observations partitioned by sufficiency of sample size (i.e., focal MDC count) for a reliable bootstrap result.
Panel \ref{fig:vanilla-summary-results:fit-effect-big} shows screen results for small effect sizes with a larger-scale dataset.
Significance thresholds marked at $p<0.05$,  $p<0.01$, and $p<0.001$, and indicated by color coding.
Sben/Gdel conditions correspond to true multilevel selection dynamics, tested across weak ($1.1\times$), moderate ($1.3\times$), and strong ($2\times$) effect sizes.
Comparator sets were binned by simulation month.
}
\label{fig:vanilla-summary-results}
\end{figure*}

Given the promising results from example screens shown in Figure \ref{fig:results-vanilla-example}, we next sought to assess detection outcomes across replicate experiments and smaller effect $1.1\times$ and $1.3\times$ effect sizes.
Figure \ref{fig:vanilla-summary-results} summarizes bootstrap tests for depression of mean normalized clade size and duration associated with focal MDCs.
Experiments were conducted under baseline ``vanilla'' configuration.

%Excluding replicates with fewer than 30 observed MDCs (due to bootstrap unreliability for small sample sizes),
0/5 replicates with true neutral focal sites screened positive at $\alpha=0.05$ (although all 5 of these replicates had too few observed MDCs for a reliable bootstrap analysis).
Across the tested Sben/Gneu effect sizes, 1/15 focal sites screened positive. However, this replicate had fewer than 30 observed MDCs, making the bootstrap analysis unreliable.
Under true multilevel selection conditions (Sben/Gdel), we detected clade fitness statistics below the 50th percentile for all five $2\times$ effect size replicates.
However, sensitivity was insufficient to reliably detect smaller $1.1\times$ and $1.3\times$ effect sizes, with only 2/10 replicates screening positive.

\subsection{Fitness Effect Screen: Dataset Size}

To assess the extent to which assay sensitivity could be improved by increases in dataset size, we performed additional replicates with a larger simulated population of 1.2 million agents.
Under these conditions, we were able to reliably detect true multilevel selection dynamics in 5/5 trials at the $1.3\times$ effect size treatment.
However, no multilevel selection effect was detected at the $1.1\times$ effect size with this larger population size.
Specificity remained fair for these larger population size trials, with false positive detections under Sben/Gneu conditions for 0/10 trials.
Figure \ref{fig:vanilla-summary-results:fit-effect-big} shows these results.

\subsection{Fitness Effect Screen: Environmental Fluctuations}

We next assessed the robustness of comparator-based screens for negative MDC-associated fitness effects over populations under the influence of extrinsic, time-varying environmental factors.
For these experiments, we utilized Covasim configurations modeling vaccine rollout and public health policies over the course of the pre-omicron COVID-19 pandemic.
Due to significantly lower case counts within this configuration, we conducted all trials at the larger 1.2 million host population size in order to achieve a pathogen phylogeny dataset size comparable to baseline conditions.
Under these conditions, we found comparable sensitivity, with true positive detections in 5/5 trials under $2\times$ effect size.
However, for smaller, $1.1\times$ and $1.3\times$ effect sizes, true multilevel selection dynamics (Sben/Gdel) were only detected in 1/10 replicates (and the number of MDCs was too low for bootstrap reliability).
We also found specificity to be comparable to baseline conditions, with false positive detections in 3/20 trials (all with too few MDCs for reliability).
Supplementary Figure \ref{fig:uk-summary-results} summarizes these results.

\subsection{Fitness Effect Screen: Adaptive Dynamics}

Finally, we assessed the robustness of comparator-based screens for negative fitness effects in the presence of strong background inter-clade fitness differentials.
For these experiments, we extended our baseline Covasim configuration to incorporate introduction of alpha, beta, and gamma strains over the course of the pandemic.
In addition to partitioning MDC comparator sets by month, as was performed for other assays, for these trials we additionally assessed further partitioning by strain ID (e.g., wildtype, alpha, beta, etc.).
We include these strain-partitioned results in Supplementary Figure \ref{fig:multistrain-summary-results}, although we did not find appreciable effects of this further clade-based partitioning.

As before, under month-by-month partitioning of comparator sets, we found reliable sensitivity under a strong $2\times$ effect size.
Again, though, at smaller, $1.1\times$ and $1.3\times$ effect sizes, true multilevel selection dynamics (Sben/Gdel) were detected in only 4/10 trials.
False positive detections occurred in 2/20 trials.

\subsection{Mutation Frequency Screen}

\begin{figure}
\includegraphics[width=\linewidth]{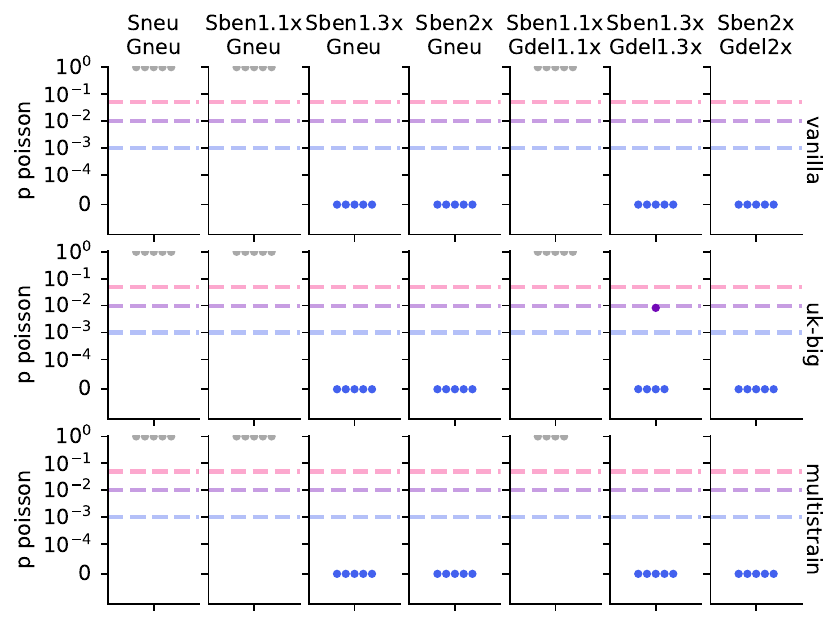}

\caption{%
\textbf{Mutation frequency screen results}.
\footnotesize
Dots correspond to screen results comparing MDC count for focal site to background sites in one replicate trial.
Significance thresholds marked $p<0.05$,  $p<0.01$, and $p<0.01$, and indicated by color coding.
Results shown for baseline (``vanilla''), varying-environment (``uk-big''), and adaptive evolution (``multistrain'') trials.
Sben/Gdel conditions correspond to true multilevel selection dynamics, tested across weak ($1.1\times$), moderate ($1.3\times$), and strong ($2\times$) effect sizes.
Positive results expected for all Sben treatments.
}
\label{fig:mutfreq-screen}
\vspace{-2ex}
\end{figure}

Figure \ref{fig:mutfreq-screen} shows focal mutation frequency detection outcomes across surveyed baseline, UK, and multistrain Covasim configurations.
Across all conditions, Sben treatments are reliably detected in all instances for $1.3\times$ and $2\times$ effect sizes.
In addition, no false-positive Sneu treatments are detected.
However, we do not find this test to be sensitive at the $1.1\times$ effect size level, with no Sben treatments detected.

Supplementary Figure \ref{fig:mutfreq-screen-counts} provides a raw view of focal MDC mutation counts observed across trials.
Notably, these counts help explain low sensitivity for small $1.1\times$ effect sizes, as actual MDC counts are highly overlapping with MDC counts under true neutral conditions.

\section{Conclusion} \label{sec:conclusion}

In this work, we have evaluated a quantitative methodology for characterizing multilevel selection effects within an evolving population and extended it to be more applicable to Artificial Life data.
% The goal of this work is to validate the sensitivity and accuracy of this method and identify potential limiting factors with respect to the scenarios in which it can be employed.
The core approach we investigated was proposed by \citet{BonettiFranceschi2024phylogenetic}.
It attempts to detect the presence of multilevel selection by identifying specific genome sites subject to it.
Although the evidence generated by this approach is circumstantial rather than directly conclusive, it is nonetheless powerful in identifying candidate loci for coupled follow-up knockout experiments.

In preliminary work, we encountered obstacles to using this approach to identify MLS in our digital system.
These obstacles stemmed primarily from 1) topological artifacts related to having a perfectly tracked (rather than inferred) phylogeny, and 2) insufficient statistical power due to the use of a single clade as the comparator.
However, after modifying the comparator normalization approach, we were able to accurately identify genome sites experiencing MLS.
This modified metric appeared to be robust to additional model complexity (with the caveat that some conditions pushed the number of comparator clades too low for the bootstrap statistical test to be viable).

Several further considerations in applying this approach to artificial study systems should be noted.
In very unstable or fast-evolving populations, a consensus genetic background for categorizing mutations may not be well-defined.
However, because ALife systems often enable direct observation of mutational events, this issue may be moot.
Finally, given broad experimental control over configuration of digital experiments, consideration should be given to tuning an elevated mutation rate during the course of data collection to enhance screen power by ensuring large MDC sample counts.

Overall, our modified method appears to be a promising tool for detecting MLS in artificial life systems.
We are optimistic about its utility for quantifying aspects of open-ended evolution, and anticipate future work evaluating and applying this approach in more sophisticated model systems used to study major evolutionary transitions.

% The second approach is based on mathematical formalism of the statistical effects of interactions between selection effects and changes in population composition.
% Under this framing, the multilevel selection dynamics can be connected to expected structural artifacts in phylogeny structure within groups versus between groups.
% Notably, unlike the Francesci and Volz approach, this method is agnostic to sequence data.

% Both approaches require phylogeny data.
% This approach requires phylogeny data accompanied by sequence data.
% Sufficient population size and sampled tip count is necessary in order to observe multiple independent originations of the same mutation.
% Obtaining this data may become difficult in scenarios where mutation rates are very low or the quantity of candidate genome sites is very high (although perhaps systematic assays injecting mutations at an elevated rate within a simulated population could be devised).

% In the Price Equation approach, the phylogeny data must include comparator taxa sampled from within the same group, as well as explicit definitions of the hypothesized groups.
% In future work, it would be interesting to explore clustering methods to hypothesize groups to allow this method to be applied to unlabeled data.

\subsection{Future Work}

% Multilevel selection is of core interest within artificial life research.
% What constitutes an organism is a fundamental question within artificial life and one that the field is uniquely suited to investigate.
% For some experiments, it is advantageous to study individuality in terms of phenomena that arise de novo within a simulation --- and therefore must be detected --- rather than being coded in a priori.
% Transitions in individuality have also been highlighted with respect to open-ended evolution.
% Again, within this framing, the key question of interest is with regard to phenomena that cannot be predicted a priori and therefore can be difficult to recognize and quantify.
% For this purpose, assessment of open-ended properties of systems can be made much more rigorous by incorporating the methodologies assessed in this work.
% In particular, given the fundamental role of replication within evolutionary dynamics, phylogeny-based methods are quite generic and can be readily generalized over a broad swath of systems --- although there may be some challenges for truly low-level systems where no fundamental replicator unit is defined a priori and therefore must be discerned on-the-fly.

In addition to the aforementioned open-ended evolution opportunities, these results suggest a number of directions for future work.
First, further comparison of Francesci and Volz's original method to our extended version would be worthwhile.
In particular, it will be important to test both methods in a digital system with known ground-truth that is designed to more closely mirror real-world phylogeny reconstruction.
Such a test will clarify whether encountered issues are specific to applying this methodology to ALife data.

More generally, this work underscores a need for high-throughput tools to collect and study phylogenies.
In conjunction with efforts to facilitate convenient, scalable recording of phylogenies from digital evolution experiments \citep{dolson2024phylotrack,lalejini2019data,moreno2022hstrat,godin-duboisAPOGeTAutomatedPhylogeny2024}, analyzing the immense quantities of resulting data requires optimized software for conducting phylogenetic analyses \citep{moshiriCompactTreeLightweightHeaderonly2025}.
As such, strong potential exists to broaden collaboration with the larger bioinformatics community in shared open-source software tools.

% In service of present work, we have utilized phyloanalysis tools built around the alife data standard, and uniting it with the powerful high-throughput dataframe ecosystem of existing tools.
% In addition to exploring cutting-edge computational approximations transformed analysis that took upwards of 40 hours to one that can be completed on the order of minutes.
% In other work, we hace scaled this approach for processing applications on billion tip phylogenies, and there is strong potential to develop this as a broader part of the high-performance phylogenetics ecosustem.
% As pursued in present work, there is also substantial potential for there to be collaborations exploiting the capability of artificial life systems to generate rich ground truth data as a testbed for bioinformatics statistics, as was pursued in present work.

\section*{Acknowledgment}
\begin{footnotesize}
Thank you to Jeet Sukumaran for insightful discussions that contributed to the development of this project.

Computational resources were provided by the MSU Institute for Cyber-Enabled Research.

This material is based upon work supported by the U.S. Department of Energy, Office of Science, Office of Advanced Scientific Computing Research (ASCR), under Award Number DE-SC0025634.
This report was prepared as an account of work sponsored by an agency of the United States Government.
Neither the United States Government nor any agency thereof, nor any of their employees, makes any warranty, express or implied, or assumes any legal liability or responsibility for the accuracy, completeness, or usefulness of any information, apparatus, product, or process disclosed, or represents that its use would not infringe privately owned rights.
Reference herein to any specific commercial product, process, or service by trade name, trademark, manufacturer, or otherwise does not necessarily constitute or imply its endorsement, recommendation, or favoring by the United States Government or any agency thereof.
The views and opinions of authors expressed herein do not necessarily state or reflect those of the United States Government or any agency thereof.

This material is based upon work supported by the Eric and Wendy Schmidt AI in Science Postdoctoral Fellowship, a Schmidt Sciences program.
\end{footnotesize}

\putbib

\end{bibunit}

\clearpage
\newpage

\begin{bibunit}

\appendix
\onecolumn

\section*{\Huge Supplemental Material for ``Extending a Phylogeny-based Method for Detecting Signatures of Multi-level Selection for Applications in Artificial Life''}

\FloatBarrier

\setcounter{section}{0}

% https://tex.stackexchange.com/a/708552/316176
\makeatletter
\def\@seccntformat#1{\@ifundefined{#1@cntformat}%
   {\csname the#1\endcsname\space}%    default
   {\csname #1@cntformat\endcsname}}%  enable individual control
\newcommand\section@cntformat{\thesection.\space} % section-level
\makeatother
\renewcommand{\thesection}{S\arabic{section}}
\counterwithin{equation}{section}
\counterwithin{figure}{section}
\counterwithin{table}{section}
\counterwithin{theorem}{section}
\counterwithin{algorithm}{section}
\counterwithin{lstlisting}{section}

\begin{figure*}
\includegraphics[width=\linewidth]{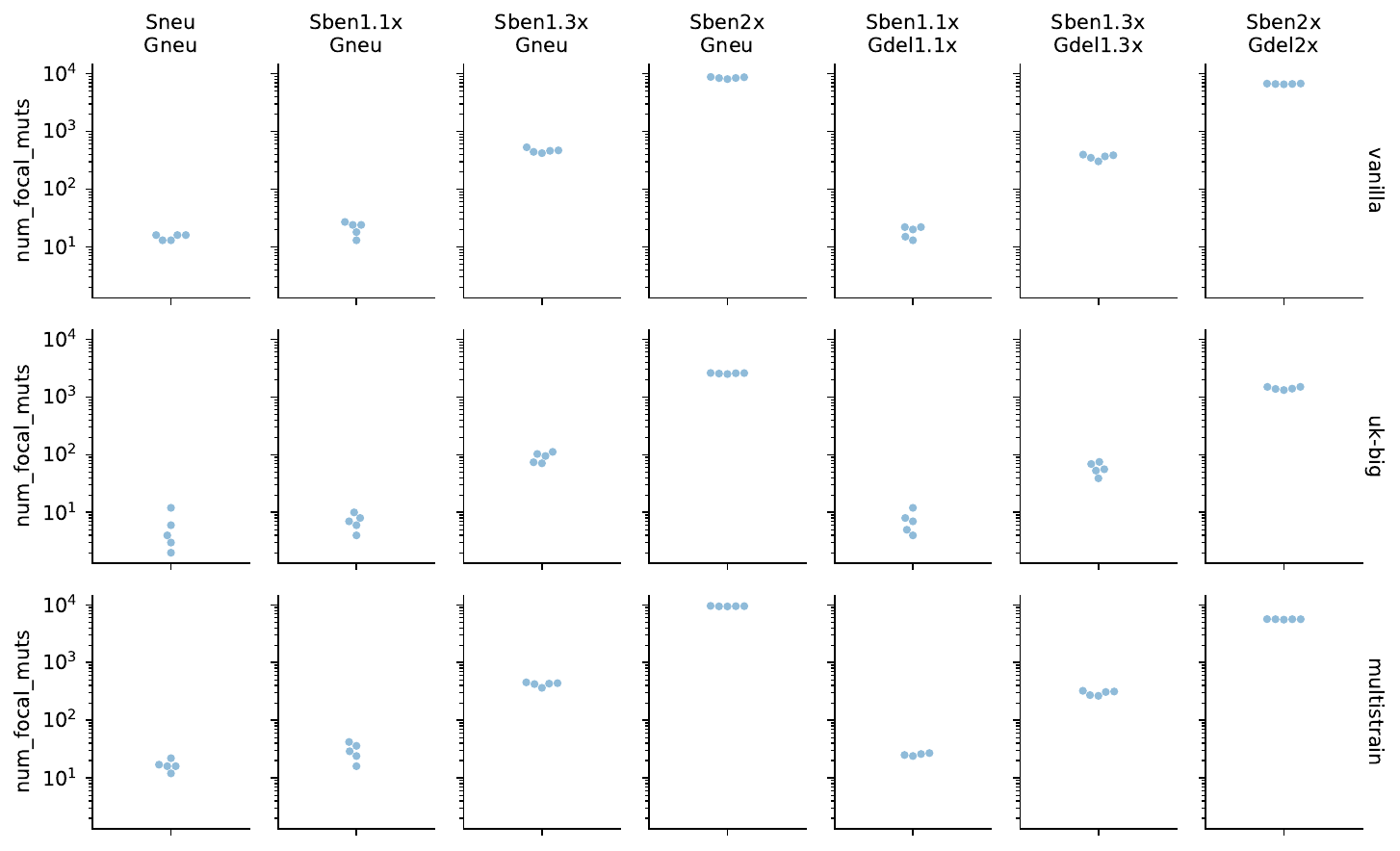}

\caption{%
\textbf{Focal-site mutation-defined clade counts.}
\footnotesize
Dots indicate the number of focal site MDCs observed per replicate trial.
Note log-scale $y$ axis.
Results shown for baseline (``vanilla''), varying-environment (``uk-big''), and adaptive evolution (``multistrain'') trials.
}
\label{fig:mutfreq-screen-counts}
\end{figure*}

\begin{figure*}
\centering

\begin{minipage}{0.32\linewidth}
\includegraphics[height=\linewidth, angle=90, trim={2cm 0 2.5cm 2cm}, clip]{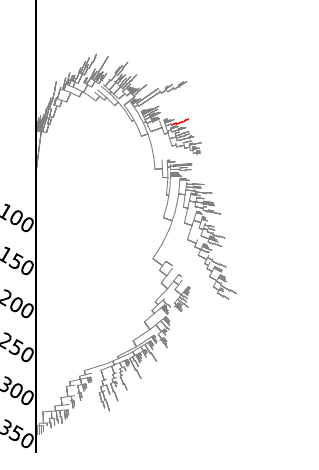}
\subcaption{Sneu/Gneu}
\label{fig:phylos-vanilla-example:sben-gneu}
\end{minipage}

\begin{minipage}{0.32\linewidth}
\includegraphics[height=\linewidth, angle=90, trim={2cm 0 2.5cm 2cm}, clip]{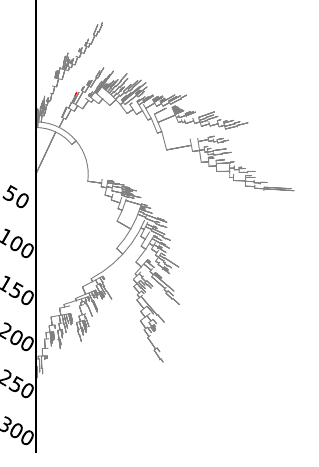}
\subcaption{Sben1.1$\times$/Gneu}
\label{fig:phylos-vanilla-example:sben1.1x-gneu}
\end{minipage}
\begin{minipage}{0.32\linewidth}
\includegraphics[height=\linewidth, angle=90, trim={2cm 0 2.5cm 2cm}, clip]{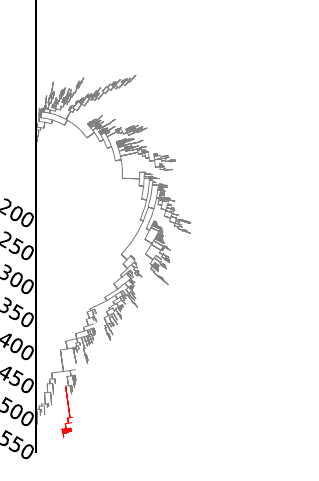}
\subcaption{Sben1.3$\times$/Gneu}
\label{fig:phylos-vanilla-example:sben1.3x-gneu}
\end{minipage}
\begin{minipage}{0.32\linewidth}
\includegraphics[height=\linewidth, angle=90, trim={2cm 0 2.5cm 2cm}, clip]{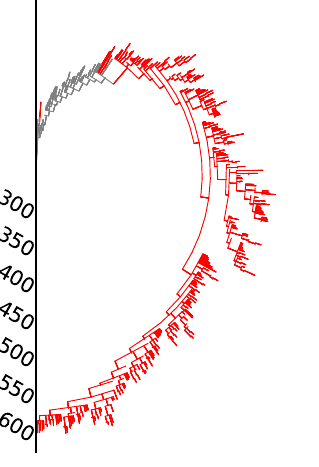}
\subcaption{Sben2$\times$/Gneu}
\label{fig:phylos-vanilla-example:sben2x-gneu}
\end{minipage}

\begin{minipage}{0.32\linewidth}
\includegraphics[height=\linewidth, angle=90, trim={2cm 0 2.5cm 2cm}, clip]{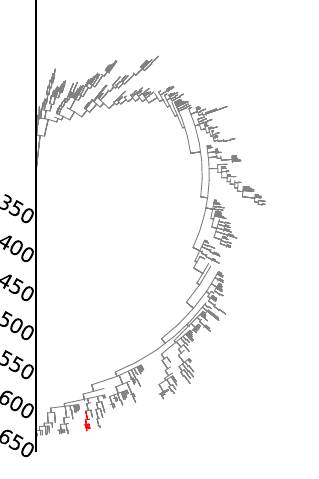}
\subcaption{Sben1.1$\times$/Gdel1.1$\times$}
\label{fig:phylos-vanilla-example:sben1.1x-gdel1.1x}
\end{minipage}
\begin{minipage}{0.32\linewidth}
\includegraphics[height=\linewidth, angle=90, trim={2cm 0 2.5cm 2cm}, clip]{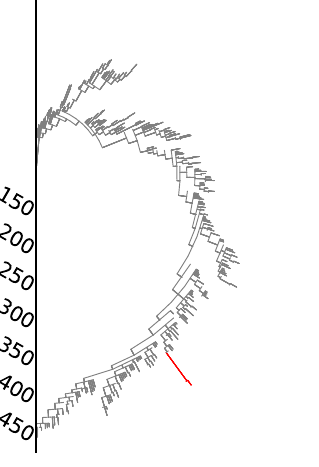}
\subcaption{Sben1.3$\times$/Gdel1.3$\times$}
\label{fig:phylos-vanilla-example:sben1.3x-gdel1.3x}
\end{minipage}
\begin{minipage}{0.32\linewidth}
\includegraphics[height=\linewidth, angle=90, trim={2cm 0 2.5cm 2cm}, clip]{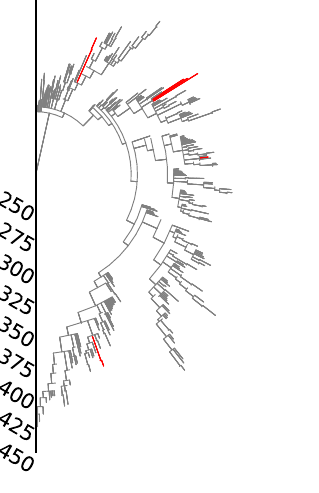}
\subcaption{Sben2$\times$/Gdel2$\times$}
\label{fig:phylos-vanilla-example:sben2x-gdel2x}
\end{minipage}

\caption{%
\textbf{Representative phylogenies from vanilla conditions.}
\footnotesize
For legibility, a sampled subclade from one replicate per treatment is illustrated.
Subclades sized between 500 and 750 taxa containing at least one instance of the variant of interest were chosen.
Trees are time-scaled with respect to days elapsed.
Red branches are the variant of interest.
}
\label{fig:phylos-vanilla-example}

\end{figure*}

\begin{figure*}
\centering
\begin{minipage}{\linewidth}
\includegraphics[width=\linewidth]{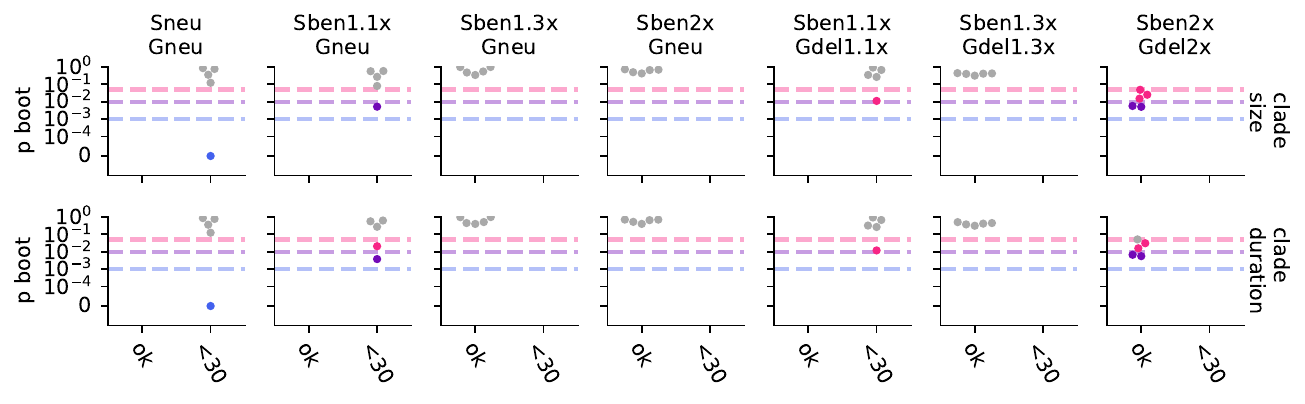}
% \subcaption{fitness effect screen covasim host population size 1.2 million}
% \label{fig:uk-summary-results:fit-effect-small}
\end{minipage}

\caption{%
\textbf{Fitness effect screen results for varying-environment trial.}
\footnotesize
Dots correspond to screen results for focal site in one replicate trial, conducted under configuration for UK epidemiological interventions with host population size 1.2 million.
Observations partitioned by sufficiency of sample size (i.e., focal MDC count) for a reliable bootstrap result.
Significance thresholds marked at $p<0.05$,  $p<0.01$, and $p<0.001$, and indicated by color coding.
Sben/Gdel treatment corresponds to true multilevel selection dynamics.
Comparator sets were binned by simulation month.
}
\label{fig:uk-summary-results}
\end{figure*}

\begin{figure*}
\centering
\begin{minipage}{\linewidth}
\includegraphics[width=\linewidth]{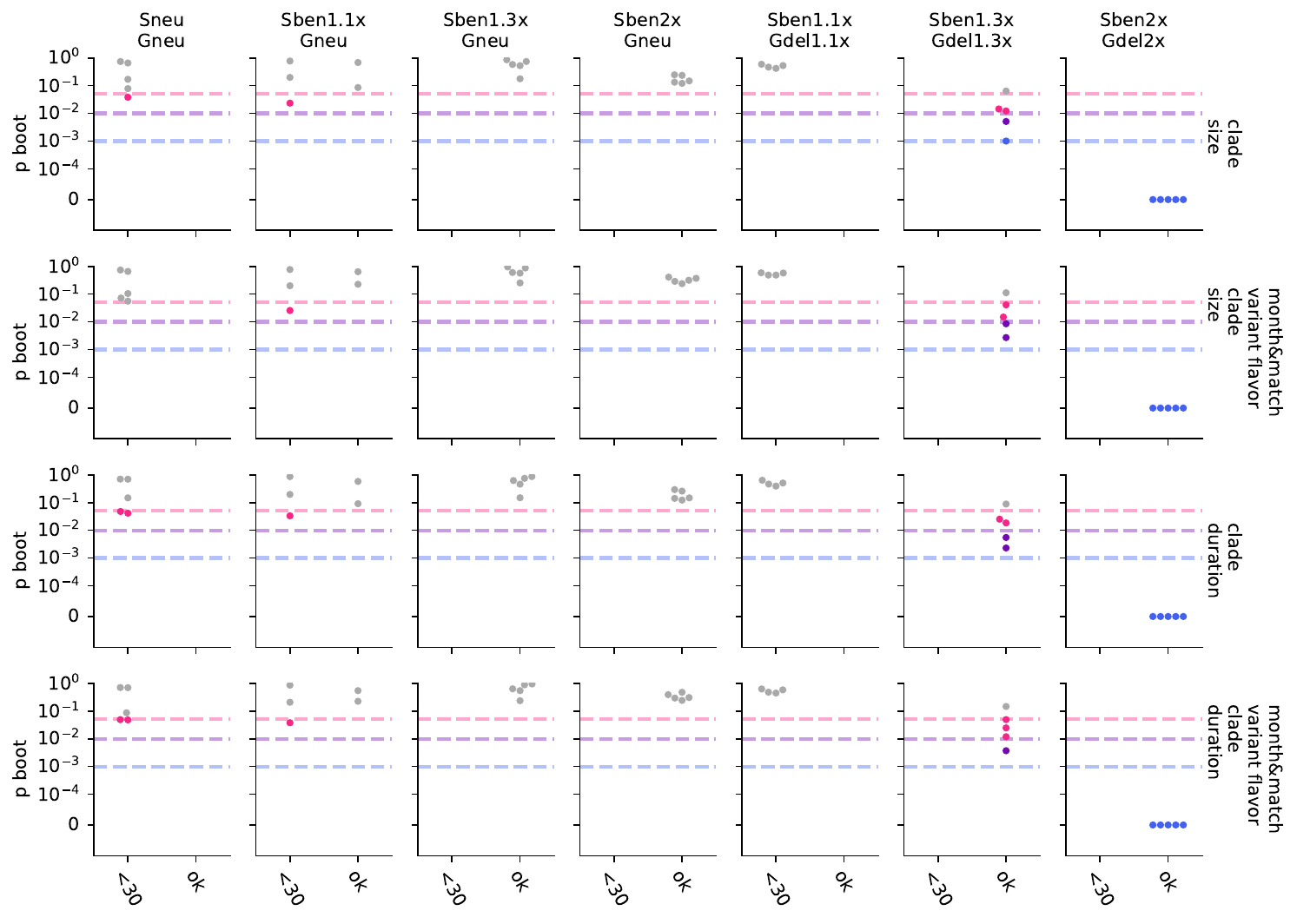}
\end{minipage}

\caption{%
\textbf{Fitness effect screen results for multistrain trial.}
\footnotesize
Dots correspond to screen results for focal site in one replicate trial, conducted with underlying adaptive dynamics.
Observations partitioned by sufficiency of sample size (i.e., focal MDC count) for a reliable bootstrap result.
Significance thresholds marked at $p<0.05$,  $p<0.01$, and $p<0.001$, and indicated by color coding.
Sben/Gdel treatment corresponds to true multilevel selection dynamics.
Comparator sets were binned by simulation month (rows 1 and 3) or by simulation month and variant (rows 2 and 4).
}
\label{fig:multistrain-summary-results}
\end{figure*}

\subsection{Simulation Dynamics}

% vanilla logs https://osf.io/tmg6b
% https://github.com/mmore500/multilevel-selection-concept/blob/2db19e1426e09da06006252ca82708a5bf5935ac/binder/2025-05-30-parse-logs.ipynb
% vanilla mdc counts https://github.com/mmore500/multilevel-selection-concept/blob/62d8f1eff044a31b89c80316b03585613661d952/binder/2025-05-19-compscreen-vanilla-summary.ipynb
% mut counts https://github.com/mmore500/multilevel-selection-concept/blob/355809574ce6bacec432b33cb6133b9658ccfc41/binder/2025-05-30-compscreen-mutcount.ipynb
Under baseline conditions, net case counts of 709,268 (SD 9,553) infections were observed over the two-year simulation window among the population of 200,000 agents.
Under Sneu/Gneu treatment conditions, for the focal mutation of interest on average 14.8 (SD 4.2) MDCs comprising 85 (SD 81) mutant samples were observed.
For Sben/Gneu treatments, 21.2 (SD 5.6)/466 (SD 41)/8,493 (SD 284) focal MDCs were observed across effect size strengths, comprising 525 (SD 479)/29,040 (SD 10,708)/299,979 (SD 19,759) total focal mutants.
For Sben/Gdel treatments, 18.4 (SD 4.2)/361 (SD 37)/6,697 (SD 82) focal MDCs were observed across effect size strengths, comprising 141 (SD 104)/6,154 (SD 9,340)/18,099 (SD 6,142) total focal mutants.
Supplementary Figure \ref{fig:phylos-vanilla-example} shows sample pathogen phylogenies observed in baseline condition experiments.

% uk logs https://osf.io/ej5bz
% uk big logs https://osf.io/nm6wz
% multistrain logs https://osf.io/qdwb7
% https://github.com/mmore500/multilevel-selection-concept/blob/2db19e1426e09da06006252ca82708a5bf5935ac/binder/2025-05-30-parse-logs.ipynb
Under baseline conditions with a population of 1.2 million host agents, mean net case counts were 4,212,074 (SD 15,273) infections
Under configurations modeling UK vaccination rollout and public health measures, case counts were much lower --- on average, 18,902 (SD 4,857) infections for population size 200,000.
For population size 1.2 million, UK-configured mean case counts were 93,412 (SD 13,418) infections.
Likewise, with the introduction of high-transmissibility strains in our multistrain experiments, case counts were higher than under baseline conditions, on average 1,212,595 (SD 5,645) infections for a population size of 200,000 hosts.

Focal mutant and focal MDC counts across all surveyed scenarios are provided at \url{https://osf.io/f6372}.

For analysis pipeline tractability, phylogenies were downsampled to at most 1 million tips for 200,000-agent configurations and at most 2 million tips for 1.2 million-agent configurations.

\end{bibunit}

\end{document}